\documentclass{article}
\usepackage{spconf,amsmath,graphicx}
\usepackage{booktabs}
\usepackage{microtype}
\usepackage{subfigure}
\usepackage{gensymb}
\usepackage{multirow}
\usepackage{xcolor}
\usepackage{makecell}
\usepackage[section]{placeins}
\usepackage{textcomp}
\usepackage{url}

\renewcommand{\paragraph}[1]{
     \noindent\textbf{#1:} 
 }

 \newcommand\blfootnote[1]{%
  \begingroup
  \renewcommand\thefootnote{}\footnote{#1}%
  \addtocounter{footnote}{-1}%
  \endgroup
}
\newcommand{\gsound}{GAS}

\DeclareMathOperator{\boldx}{\mathbf{x}}

\graphicspath{ {figures/} }

\title{Improving Reverberant Speech Training Using \\Diffuse Acoustic Simulation}
\name{Zhenyu Tang$^{\star}$ \qquad Lianwu Chen$^{\dagger}$ \qquad Bo Wu$^{\dagger}$ \qquad Dong Yu$^{\dagger}$ \quad Dinesh Manocha$^{\star}$}

			\address{$^{\star}$ University of Maryland \quad $^{\dagger}$ Tencent AI Lab\\
			    \{zhy, dm\}@cs.umd.edu, \{lianwuchen, lambowu, dyu\}@tencent.com}
\begin{document}
%\ninept
%
\maketitle
\begin{abstract}
We present an efficient and realistic geometric acoustic simulation approach for generating and augmenting training data in speech-related machine learning tasks. Our physically-based acoustic simulation method is capable of modeling occlusion, specular and diffuse reflections of sound in complicated acoustic environments, whereas the classical image method can only model specular reflections in simple room settings. We show that by using our synthetic training data, the same neural networks gain significant performance improvement on real test sets in far-field speech recognition by 1.58\% and keyword spotting by 21\%, without fine-tuning using real impulse responses.

\end{abstract}
\begin{keywords}
reverberation, diffuse reflection, speech recognition, data augmentation, acoustic simulation
\end{keywords}

\section{Introduction}
Over the past few years, deep learning approaches have gained significant ground in the speech community, surpassing the performance of many classical machine learning models in a variety of related sub-fields. State-of-the-art deep neural networks (DNNs) are powerful tools for exploiting variable-length contextual information embedded in noisy speech sequences. \blfootnote{This work is supported in part by ARO grant W911NF-18-1-0313, NSF grant \#1910940, Tencent, Adobe, Facebook and Intel. The authors thank Jie Chen and Dan Su from Tencent for their help with the ASR and KWS systems. Project website \url{https://gamma.umd.edu/pro/speech/asr}}
Some very famous applications of DNN techniques in speech include Microsoft Cortana\textregistered, Apple Siri\textregistered, Google Now\textregistered, and Amazon Alexa\textregistered. These applications usually integrate several fundamental speech tasks such as speech enhancement and separation~\cite{hershey2016deep,yu2017permutation}, automated speech recognition (ASR)~\cite{seide2011conversational,dahl2011context, Xiong2017The,yu2017recent}, and keyword spotting (KWS)~\cite{GuoguoChen2014,Prabhavalkar2015Automatic}.
%and speaker identification and recognition~\cite{Yun2014A,zhang2017end}. 
Another important enabling factor behind the success of DNNs in these tasks is the huge amount of annotated speech corpus made available by research groups and large companies. Deep learning theory indicates that having more training examples is crucial to reduce the generalization error of trained models in real test cases \cite{Seltzer2013An}. However, the majority of popular speech corpuses were recorded under relatively ideal conditions, i.e. anechoic speech with negligible noise and environmental reverberation. When training models for real-world applications, it is common to distort the clean speech by adding noise and reverberation as a pre-processing step to augment the training data \cite{Kim2017Generation,Doulaty2017Automatic}. Reverberation is a characteristic effect of a particular acoustic environment and can be described by impulse responses (IRs) or frequency responses. In practice, both recorded IRs and synthetic IRs have been used to convolve with the clean speech. Significant improvements in model accuracy have been observed due to this type of data augmentation. However, there is still a performance gap when the application is deployed in conditions not matched to training conditions. IRs pre-recorded in a limited number of environments may not generalize well to infinite real-world conditions. However, it is in-efficient to shrink the gap by collecting more real-world IRs; recording IRs is not a trivial task because it requires professional equipment and trained people. An alternative and cost-effective way is simulating room impulse responses (RIRs) by using acoustic simulators. A simple RIR simulator should take in the room geometry, source and listener positions, and surface absorption/reflection properties, and generate an RIR for each source-listener pair. One classical approach is the image method (IM), which models specular sound reflections in rectangular rooms and has been proven to work well in some tasks. However, one notable drawback of this method is its over-simplification of room acoustics by ignoring diffuse reflections that are very common in real-world environments. Furthermore, it does not deal with occlusion. These limitations make the image method less realistic in terms of augmenting data, especially in applications where late reverberation plays a significant role. 

\paragraph{Main contribution} 
To overcome limitations of existing simulation methods and better augment the training data, we propose an efficient and realistic geometric acoustic simulation approach that models occlusion, specular, and diffuse reflections, where sound energy can be reflected randomly and thus not following an ideal specular path. We sample 5000 different acoustic room configurations and use our method to simulate far-field sound propagation in each room. The speech training data is generated by randomly convolving over 1500 hours of clean speech utterances with simulated RIRs and adding environmental noise. We train two different models independently based on 1D/2D convolution + long short-term memory (LSTM) structures for an ASR task and a KWS task, and then evaluate them on different real-world data. We observe accuracy improvement using our method by 1.58\% in terms of ASR and by 21\% in terms of KWS.

The rest of the paper is organized as follows. In Section 2 we explain our ray-tracing based geometric acoustic simulation algorithm. We describe several speech training benchmarks in Section 3 and present our results in Section 4. 

\section{Acoustic Simulation}
% \chlw{probably we need to give a brief introduction of the image method in this section, since some people in ASR area are not familiar with it, like \cite{Kim2017Generation}. Besides, it would be better if we have some figures to illustrate the difference of the image method and the geometric acoustic simulation method in this section or section 4, especially for the diffuse reflections.}

\subsection{Impulse Response Modeling}
Acoustic simulation engines have been used in computer aided designs (CAD), theoretical research, the game industry, and many other fields. The simulation goal is usually to observe how the sound pressure changes according to time at some position when there is a sound source at some other position in space. IR is the most common way to describe sound propagation between two points in a fixed environment, so we use $\text{IR}(\boldx_s,\boldx_r,t)$ to denote the IR at time $t$ from the point source at location $\boldx_s$ to the listener at location $\boldx_r$. In practice, an IR can be measured by exciting an impulse using a shotgun as a sound source; the sound pressure is then recorded at the target receiver location. 
% From a first principle view, the propagation of sound waves in a homogeneous medium is governed by the following acoustic wave equation:
% \begin{equation}
% \label{eq:waveequation}
% \nabla ^2 p(\boldx, t) - \frac{1}{c^2} \frac{ \partial^2 p(\boldx, t)}{\partial t ^2 } = f(\boldx, t),
% \end{equation}
% where $p(\boldx, t)$ is the acoustic pressure in \emph{Pascal} at time $t$ and location $\boldx$, $c$ is the speed of sound in this medium, and $f(\boldx, t)$ is the source distribution function. By replacing $f(\boldx, t)$ with an impulse function $\delta(\boldx-\boldy)\delta(t-\tau )$, the solution to equation~(\ref{eq:waveequation}) will be the IR of all receiver locations with respect to the source location $\boldy$. This is also the foundation of wave-based solvers. 
From a first principle view, the propagation of sound waves follow the acoustic wave equation~\cite{feynman2011feynman}, which describes the sound pressure variation in both spatial and temporal domain and is the foundation of wave-based solvers. There are several ways to implement wave-based solvers, including Finite Element Methods (FEM), Boundary Element Methods (BEM), finite-difference time domain (FDTD) approaches~\cite{sakamoto2006numerical}, and Adaptive Rectangular Decomposition (ARD) methods~\cite{raghuvanshi2009efficient}. Wave-based techniques yield the most accurate results, but are only feasible for low frequencies and small scenes because they do not scale well with space and time granularity.

When the wavelength of the sound is smaller than the size of the obstacles in the environment, the sound wave can be treated in the form of a ray, which is the key idea of geometric methods. Typical geometric methods include the image method~\cite{allen1979image}, path tracing methods \cite{taylor2009resound,taylor2012guided,schissler2016interactive,schissler2018interactive}, and beam or frustum tracing methods \cite{funkhouser1998beam,chandak2008ad}. Our method is based on efficient Monte Carlo path tracing~\cite{kajiya1986rendering}.
% They are mostly accurate for higher frequencies and can be used for interactive applications.
\vspace{-0.5em}
\begin{figure}[htbp]
    \centering
    \subfigure[Specular reflections]{
        \includegraphics[width=0.35\linewidth]{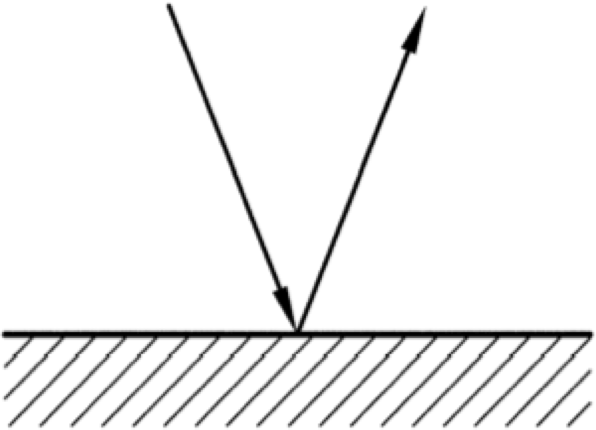}
        \label{fig:specular}
    }
    \subfigure[Diffuse reflections]{
        \includegraphics[width=0.35\linewidth]{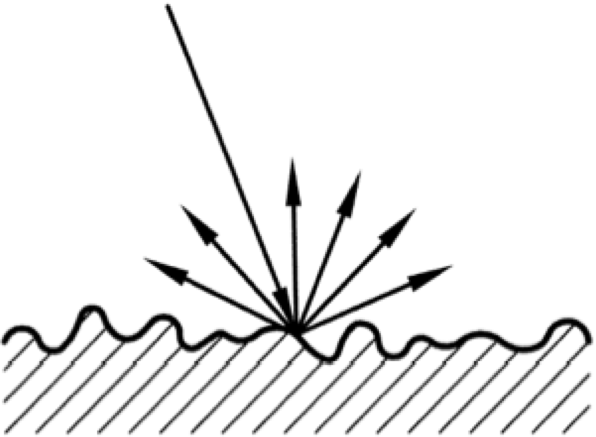}
        \label{fig:diffuse}
    }
    % \subfigure[Frequency dependent scattering]{
    %     \includegraphics[width=0.3\linewidth]{freq-dep}
    %     \label{fig:freq-dep}
    % }
    \vspace{-1em}
\caption{Two types of reflections of sound at a surface. Both phenomena are frequency dependent.}
\vspace{-1.5em}
\label{fig:sound_interaction}
\end{figure}

\subsection{Sound Propagation}
\begin{figure}[htbp]
    \centering
    \includegraphics[width=\linewidth]{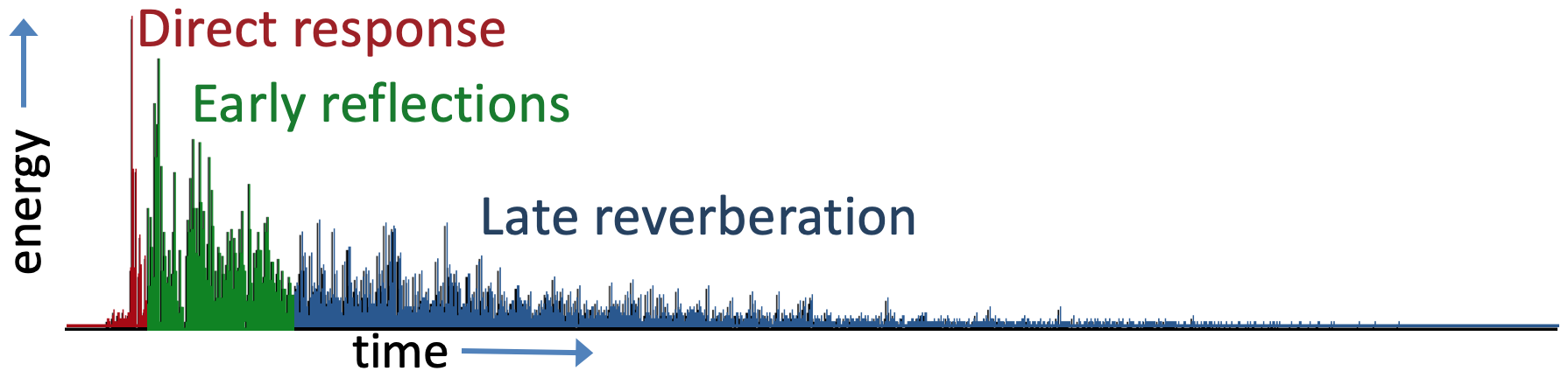}
    \vspace{-1.5em}
    \caption{Energy distribution of an impulse response in time. Our goal is to accurately model the late reverberation effects in simulated IRs.}
    \vspace{-2em}
    \label{fig:energy_dist}
\end{figure}

% \subsubsection{Specular and Diffuse Reflections}
From the perspective of geometric methods, there are two types of reflections that can occur at a rigid surface: specular reflections and diffuse reflections. Specular reflections occur at mostly flat and uniform surfaces and the outgoing direction of the sound ray is the same as the incident angle in Fig.~\ref{fig:specular}, known as Snell's Law in geometric optics. However, real-world object surfaces usually do not completely satisfy the specular condition and scatter sound energy in all directions according to Lambert's cosine-law, which is called diffuse reflections as illustrated in Fig~\ref{fig:diffuse}. IRs are constructed by accumulating sound energy from both specular and diffuse reflection paths with the correct time delay and energy decay, which can be calculated from the total length of the path. Conventionally, an IR is decomposed into 3 parts: direct response, early reflections, and late reverberation. The direct response is determined by the visibility between the source and listener. Early reflections are mostly due to specular reflections, whereas the late reverberation is caused by diffuse reflections. A typical IR energy distribution is shown in Fig.~\ref{fig:energy_dist}.

% \subsubsection{Sound Diffractions}
% Diffraction is a wave phenomenon that is distinct from specular or diffuse reflections. Sound diffraction is most noticeable when the wavelength of sound is comparable to the size of the obstacle. For example, people can hear sound from a non-line-of-sight sound source at the edge of a wall due to diffraction. Edge-based diffraction can be approximated using the Biot-Tolstoy-Medwin (BTM) model~\cite{svensson1999analytic} or the uniform theory of diffraction (UTD) approach~\cite{tsingos2001modeling}. However, these methods are limited to low complexity geometry and low order diffraction. A recent diffraction kernel method~\cite{rungta2018diffraction} can handle these difficulties with efficient precomputation. 

\subsection{Image Method}
\begin{figure}[htbp]
    \centering
    \includegraphics[width=0.5\linewidth]{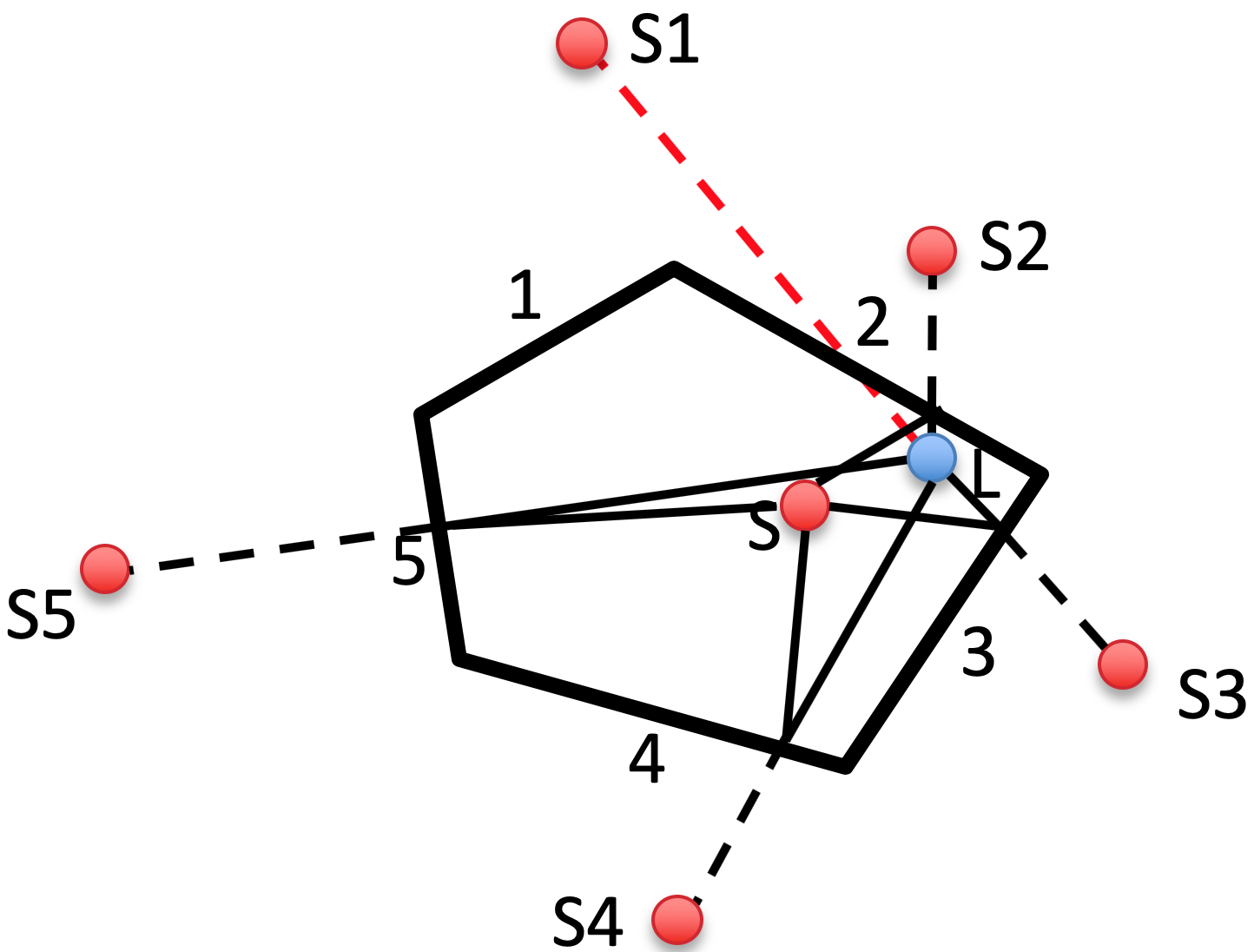}
    \vspace{-1em}
    \caption{Construction and validation of image paths. The source $S$ is mirrored into 5 image sources marked as $S1\sim S5$ by 5 planes. A sound path is connected to the listener $L$ from each image source. Then a path validation is performed by checking whether the image path intersects with the plane that generates this image source. The path from $S1$ to $L$ does not intersect with plane 1; therefore it is infeasible and rejected. The other 4 image paths are valid and can be used to compute the IR analytically.}
    \vspace{-2em}
    \label{fig:image_method}
\end{figure}

The image method is the current most widely used method in the speech community for generating RIRs in various learning-based tasks~\cite{ko2017study}. It is based on the principle of specular reflections where all reflection paths can be constructed by mirroring sound sources with respect to the reflecting plane, shown in Fig~\ref{fig:image_method}. A source will be mirrored multiple times depending on the desired order of reflections. Therefore, the image method fails to model the late reverberation part of an IR. Computationally, for a scene with one source, $N$ reflective surfaces, and reflection order $d$, the time complexity is $\mathcal{O}(N^d)$, which is prohibitive for simulations at high orders or scene complexities. 
\vspace{-1em}

\subsection{Diffuse Acoustic Simulation}
Diffuse reflections occur when sound energy is scattered into non-specular directions. Diffuse reflections are widely observed in real-world and have been shown to be important for modeling sound fields in room environments~\cite{hodgson1991evidence,dalenback1994macroscopic,tang2019scene}. Diffuse acoustic simulations correctly model not only the specular, but also the diffuse soundfield. 

\begin{figure}[htbp]
    \centering
    \includegraphics[width=\linewidth]{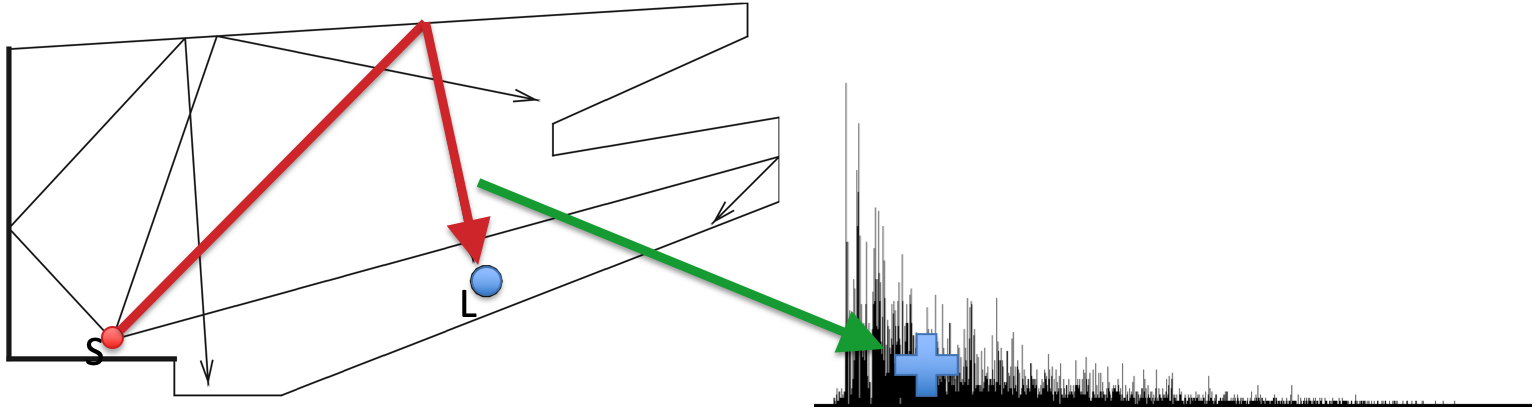}
    \vspace{-0.5em}
    \caption{Monte Carlo path tracing for solving the sound transport problem. Ray samples are generated in random directions from the source $S$. Reflections upon hitting a plane are simulated by generating subsequent random rays while conserving the total energy. Once a ray intersects with the listener $L$, the energy is accumulated to the IR.}
    \vspace{-1.5em}
    \label{fig:path_tracing}
\end{figure}

We propose our geometric acoustic simulation (\gsound) method for this purpose. In contrast to the image method, our method is based on stochastic path tracing illustrated in Fig.~\ref{fig:path_tracing}: sound paths are randomly traced in all directions and each path follows either specular or diffuse reflections. We explicitly define the scattering coefficient $s$ between 0 and 1, which denotes the proportion of sound energy that is diffusely reflected at a surface (0 means perfectly specular and 1 means perfectly diffuse). Specifically, the sound energy $L_r$ reflected at a surface point $\boldx$ to direction $\vec{\omega}_{r}$ is computed by integrating the incoming energy over a hemisphere $\Omega$ centered at $\boldx$ on the surface:
\begin{equation}
\label{eq:rendereq}
L_{r}\left(\boldx, \vec{\omega}_{r}\right)=\int_{\Omega} f_{r}\left(\boldx, \vec{\omega}_{i} \rightarrow \vec{\omega}_{r}\right) L_{i}\left(\boldx, \vec{\omega}_{i}\right) \cos \theta_{i} d \omega_{i},
\end{equation}
where $\theta$ is the incident angle, $\vec{\omega}_{i}$ is the incoming direction, and $f_{r}\left(\boldx, \vec{\omega}_{i} \rightarrow \vec{\omega}_{r}\right)$ is the probability distribution function that describes the probability of generating the sound path from $\vec{\omega}_{i}$ to $\vec{\omega}_{r}$, which is generic to include both specular and diffuse reflections. In practice, Eq.~\ref{eq:rendereq} is recursive and can only be solved numerically using Monte Carlo integration. The diffuse reflection paths are generated by tracing random rays from the source, the listener, or both~\cite{cao2016interactive}. A large number of ray samples is required for solution convergence. The complexity of Monte Carlo path tracing is $\mathcal{O}(M\text{log}N)$, where $M$ is the total number of rays traced to solve Eq.~\ref{eq:rendereq} and $N$ is the number of surfaces in the scene. One of its computational advantages over the image method is that most invalid paths that are generated, verified, and rejected in the image method are not considered in path tracing, so the number of surfaces does not greatly impact the efficiency of path tracing. This allows us to compute both early reflections and late reverberation efficiently. 

In a far-field speech simulation setting, we define an acoustic room by its length, width, and height. Acoustic absorption and scattering coefficients can be defined for each surface element (triangular mesh), which determines the relative strength of diffraction. After specifying the sound source and receiver locations within the room, our simulation generates an RIR. Detailed configurations are in Section~\ref{sec:genIR}. One speech-related problem that has benefited from more accurate simulations is the direction-of-arrival estimation task~\cite{tang2019regression}. We argue that using a more accurate geometric acoustic simulation that faithfully models the late reverberation for general speech-related training will lead to better performance in learning-based models. 

% The most common techniques for computing diffuse reflections are based on Monte Carlo path tracing,
% a probabilistic technique for solving complex integral equations (Krokstad et al., 1968; Vorlander, 1989; ¨
% Embrechts, 2000). In path tracing, rays that each carry a fraction of the total sound energy are randomly
% emitted from the source. The rays are reflected throughout the environment until an intersection with
% the listener is detected, thereby generating a sound propagation path that is accumulated to the impulse
% response. The convergence of path tracing can be improved by direct sound sampling techniques such as
% diffuse rain (Schroder et al., 2007). Path tracing is often computed in separate frequency bands to model ¨
% frequency-dependent scattering and attenuation. Radiosity may also be used for the computation of diffuse
% sound (Nosal et al., 2004), but cannot handle dynamic scenes.
\section{Training with Acoustic Simulation}
To evaluate our proposed approach, we conduct far-field automated speech recognition and keyword spotting experiments and then compare our approach with the popular image method. Both experiments are reverberant speech training tasks in which the test set is always real-world noisy reverberant speech recordings, but the training set can consist of clean speech or synthetic reverberant speech generated by either the image method or our geometric acoustic simulation.

\subsection{Impulse Response Generation}
\label{sec:genIR}
We consider a 6-microphone circular array with 7cm diameter with speakers and the microphone array randomly located in the room at least 0.3m away from the wall. Both the image method and the geometric sound simulation method were employed to simulate the impulse response randomly generated from 5000 different room configurations with the size (length-width-height) ranging from 3m-3m-2.5m to 8m-10m-6m. The distance between the speaker and microphones ranges from 0.5m to 6m. The reverberation time T60 is sampled in a range of 0.05s to 0.5s.
In general, there are two IR sets, each with 5000 IRs generated with the image method and the geometric sound simulation method, respectively. The IRs were used for data augmentation in ASR and KWS tasks. 

\subsection{Automated Speech Recognition}
\subsubsection{Data}
The training corpus consists of two sets: (i) a clean corpus of 1.5 million clean speech utterances that translates to about 1500 hours in total and (ii) a noisy far-field training set simulated based on the clean corpus by adding reverberations and mixing with various environmental noises with SNRs ranging from 0 to 24 dB. For each IR generation method, the corresponding noisy far-field training set was generated using the IRs described in Section 3.1, and the first channel of simulated data was used as the input to the ASR system. The clean speech was first used to train the acoustic model and then both the clean speech and the simulated noisy speech were used to fine-tune the model. Depending on which of the two IR simulation methods were used to generate the noisy training sets, we got two acoustic models, one for the image method and one for the GAS method. The dataset sizes for clean, the image method, and the GAS method are the same.
The testing corpus contains 2000 utterances of real far-field recording from 48 speakers; each utterance is 5 seconds on average and the whole set is about 3 hours. The data is recorded in 5 different rooms with sizes of about 4m-4m-3m. The distances between the microphones and the speaker are randomly set as 0.5 m, 1 m, 3 m and 5 m, and the SNR ranges from 5 to 20dB with the background noise of an air-conditioning or fan. 
% \vspace{-1em}
\begin{figure}[htbp]
    \centering
    \includegraphics[width=1.0\linewidth]{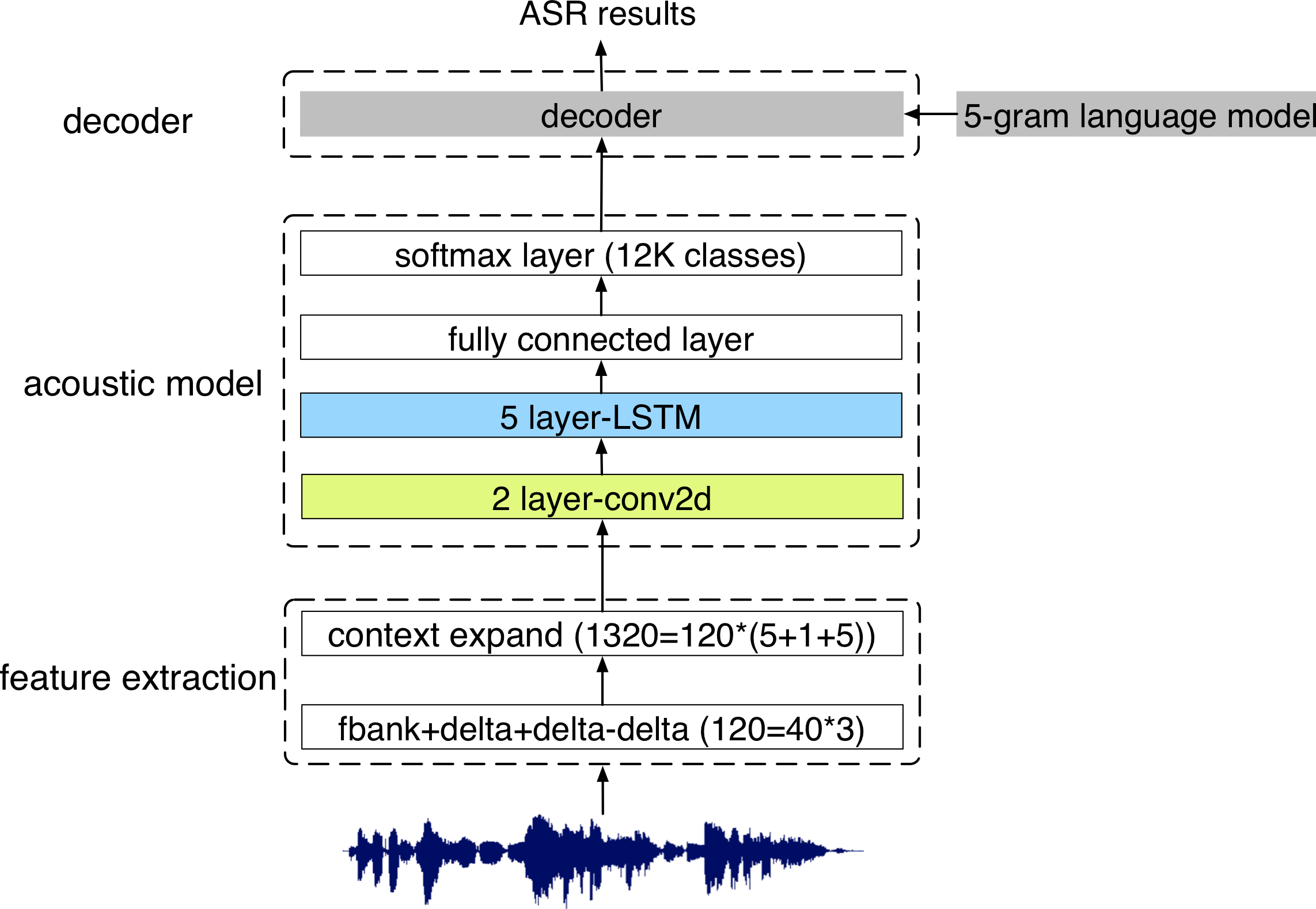}
    \vspace{-1.5em}
    \caption{The framework of our ASR system used for evaluations. }
    \vspace{-2em}
    \label{fig:acousticModel}
\end{figure}
\subsubsection{Model Configuration}
The framework of the ASR system is shown in Fig.~\ref{fig:acousticModel} and consists of feature extraction, an acoustic model ~\cite{Sainath2015Convolutional}, and a decoder. 40-dimensional Mel filter bank features were computed with a 25-ms window length and a 10-ms hop size to form a 120-dimensional vector along with their first and second order differences. After normalization, the feature vector of the current frame is concatenated with that of the 5 preceding and 5 subsequent frames, resulting in an input vector of dimension 1320 = 120 $\times$ (5 + 1 + 5). The acoustic model contains two 2-dimensional convolutional layers, each with a kernel size of (3, 3) and a stride of (1, 1), followed by a maxpooling layer with a kernel size of (2, 2) and a stride of (2, 2), and then five LSTM layers, each with 1024 hidden units and peepholes, and then one full-connection layer plus a softmax layer. Batch normalization is applied after each CNN and LSTM layer to accelerate convergence and improve model generalization. We use context-dependent (CD) phonemes as the output units, which form 12000 classes in our Chinese ASR system. The Adam optimizer was adopted with an initial learning rate of 0.0001. A 5-gram language model with size of 190 GB was used. The vocabulary's size was 280K and the training corpus was collected from news, blogs, messages, encyclopedias, etc.

\subsection{Keyword Spotting}
\subsubsection{Data}
The original training corpus contains 2500 hours of clean speech data, including 1250 hours of target keyword ``Hi, Liu Bei" and 1250 hours of negative speech samples. The corresponding multi-channel reverberant data was simulated using each IR generation method. Noises with SNRs ranging from 0 to 24dB were also added into the augmented speech. The 2500 hours of simulated reverberant data are used for model training. The test corpus contains 8000 utterances with target keyword randomly selected from real user data from smart-speakers in a typical living room scenario, as well as 33 hours of negative samples from different categories, including music, TV noise, chatter, and other indoor noises. 
The 6-channel microphone signals were processed by an MVDR beamformer~\cite{Hoshuyama2008Robust}, and the output enhanced mono-channel signal was used for keyword spotting.
\vspace{-1em}
\begin{figure}[htbp]
    \centering
    \includegraphics[width=0.8\linewidth]{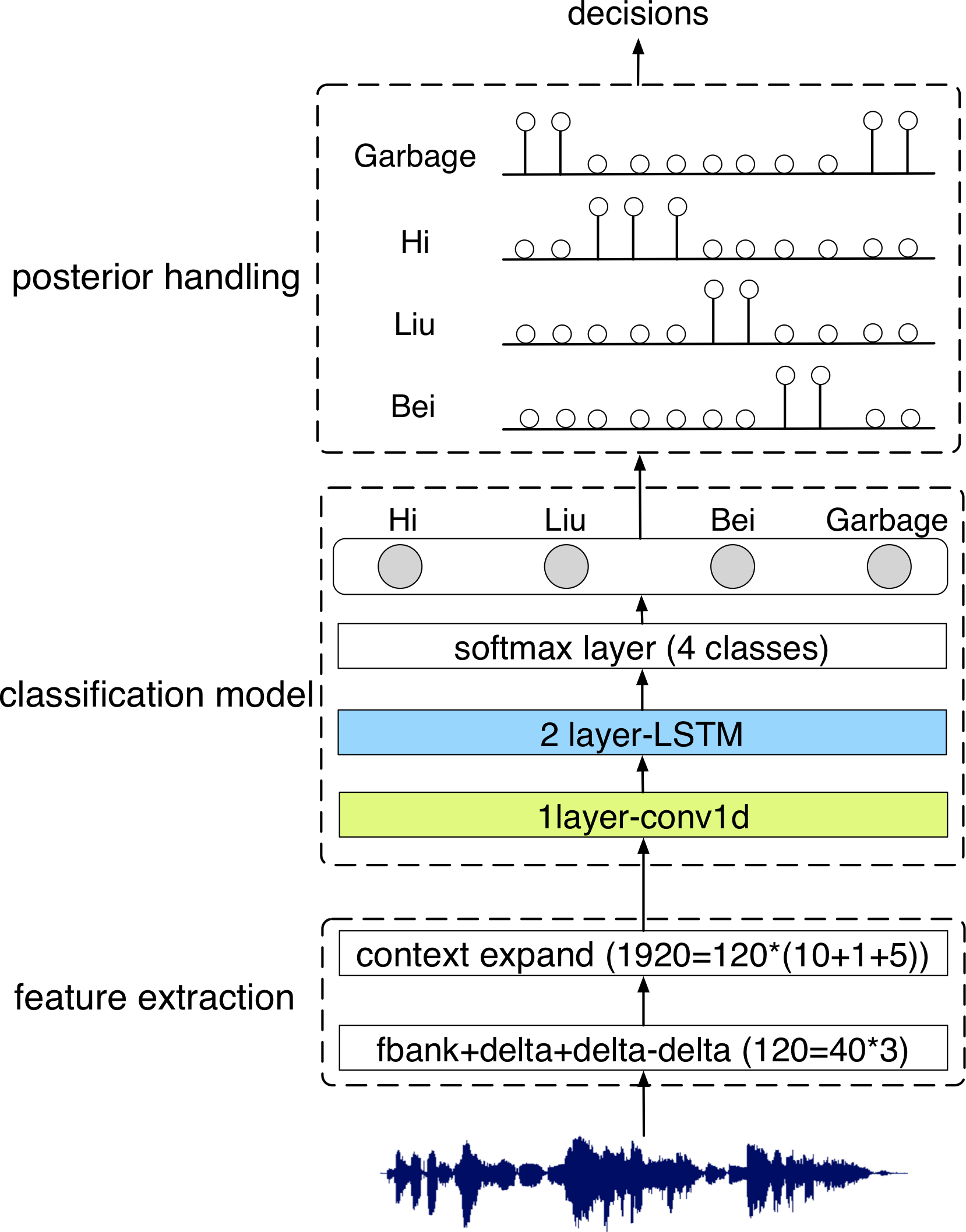}
    \caption{The framework of our KWS system used for evaluations.}
    \vspace{-1.5em}
    \label{fig:KWS}
\end{figure}
\vspace{-0.5em}
\subsubsection{Model Configuration}
 The framework of the keyword spotting system,  which is similar to ~\cite{GuoguoChen2014} is shown in Fig.~\ref{fig:KWS}, comprising feature extraction, a classification model, and a posterior handling module. The 40-dimensional Mel filter bank features were computed with a 25-ms window length and a 10-ms hop size, and then combined with the first and second order differences to form a 120-dimensional frame feature. The current frame feature was concatenated with the 10 preceding frames and 5 subsequent frames, resulting in an input vector of dimension 1920=40 $\times$ 3 $\times$ (10 + 1 + 5). The classification model contains one layer of 1D CNN~\cite{Abdel2014Convolutional} with a kernel size of 4 and is followed with a maxpooling layer with a kernel size of 3. The output of the CNN is passed to two layers of LSTM (hidden units 256) and then to a softmax layer with 4 (3 words + 1 garbage) output classes. Cross entropy is used for loss calculation. The outputs were then passed through a posterior handling module to obtain decisions. The final keyword score is defined as the largest product of the smoothed posteriors in an input sliding window, subject to the constraint that the individual words ``fire" in the same order as specified in the keyword.

\section{Results and Analysis}
\label{sec:results}
Table~\ref{tab:asr_results} shows the character accuracy of ASR systems achieved with the clean acoustic model (Clean), the noisy acoustic model based on the image method (Noisy\_IM), and the geometric sound simulation method (Noisy\_\gsound). We collected 2K real-world test utterances that are corrupted by reverberations and noises to evaluate IR methods. Compared with  the ``Clean'' setup, the ``Noisy\_IM'' setup improved the system performance significantly by adding simulated noisy training data. Our proposed approach outperformed the image method by increasing the accuracy from 59.96\% to 61.54\%, illustrating the superiority of the proposed realistic geometric sound simulation approach.
\vspace{-2em}
\begin{table}[htbp]
    \caption{Character accuracy of ASR systems. Our \gsound  method has the highest accuracy and outperforms IM by 1.58\%.}
    \centering
    \begin{tabular}{@{}lc@{}}
        \toprule
        Model  & \%  \\ \hline
        Clean & 31.178 \\
        Noisy\_IM & 59.961 \\
        Noisy\_\gsound  & \textbf{61.540}
        \\ \bottomrule
    \end{tabular}
    \label{tab:asr_results}
\end{table}

%\begin{table}[htbp]
%    \caption{Character accuracy of ASR system}
%    \centering
%    \begin{tabular}{@{}lcc@{}}
%        \toprule
%        \multirow{2}{*}{Model}  & \multicolumn{2}{c}{Test Set}  %\\\cline{2-3}
%         &\thead{character accuracy \\ on test set} & \thead{best %likely \\on development set} \\
%        \hline
%        Clean & 31.178 & 0.623 \\
%        Noisy\_IM  & 59.961 & 0.506  \\
%        Noisy\_\gsound &  61.540 & 0.515
%        \\ \bottomrule
%    \end{tabular}
%    \label{tab:asr_results}
%\end{table}
%
\vspace{-2em}
\begin{table}[htbp]
    \caption{Equal error rates of KWS systems. Our \gsound  method has the lowest equal error rate and results in a 21\% error reduction relative to that of IM.}
    \centering
    \begin{tabular}{@{}lc@{}}
        \toprule
        Model  & \%  \\ \hline
        Noisy\_IM & 1.48 \\
        Noisy\_\gsound  & \textbf{1.17} 
        \\ \bottomrule
    \end{tabular}
    \label{tab:kws_results}
\end{table}

The equal error rates (EERs) of keyword spotting systems are shown in Table~\ref{tab:kws_results}. These results indicate that we can achieve an EER of 1.17\% and 1.48\% when the augmented training data was generated using the geometric sound simulation method and the image method, respectively. This translates to a 21\% EER reduction. In these experiments, the input to the keyword spotting system is the enhanced speech from an MVDR beamformer. This indicates that the proposed IRs are robust to multichannel signal processing algorithms.

In both experiments, we carefully controlled the training and evaluation conditions, where the only difference is the RIR simulation method. Due to our faithful simulation of diffuse sound reflections, the domain gap between synthetic training data and real data is further reduced and therefore we observe significant accuracy gains.
\section{Discussion and Future Work}
In this paper, we described a geometric acoustic simulation method that simulates both the specular and the diffuse soundfields for reverberant speech training. On the speech recognition and keyword spotting tasks, we showed that the proposed approach outperformed the popular image method, where the gain is mostly attributable to the more realistic simulation of reverberation and diffuse reflections. One limitation of this work is that neither method can model low-frequency or diffraction phenomena. A partial solution would be to compensate RIRs at low-frequency bands~\cite{tang2019lowfreq}.

Although we demonstrated the efficacy of the proposed approach mainly on speech recognition and keyword spotting tasks, we believe a similar improvement on performance can be achieved on tasks such as source localization \cite{takeda2016sound}, speech separation, and the cocktail problem \cite{hershey2016deep,yu2017permutation}, all of which can benefit from data-driven techniques and are future research directions. The proposed approach is thus of wide interest, especially because it can significantly reduce the effort of collecting training data under real-usage scenarios.

% \section{Acknowledgements}
% This work is supported in part by ARO grant W911NF-18-1-0313, NSF grant \#1910940, Tencent, Adobe, Facebook and Intel. The authors would like to thank Jie Chen and Dan Su from Tencent for their help with the ASR and KWS systems.

% \pagebreak
\bibliographystyle{IEEEtran}
\bibliography{mybib}

\end{document}